\documentclass[aps,prl,twocolumn,superscriptaddress]{revtex4}
\usepackage{graphicx,bm}

\begin{document}

\title{Flat rotation curves in Chern-Simons modified gravity}

\author{Kohkichi Konno}
\email[]{konno@topology.coe.hokudai.ac.jp}
\affiliation{Department of Applied Physics, Hokkaido University,
Sapporo 060-8628, Japan}
\author{Toyoki Matsuyama}
\affiliation{Department of Physics, Nara University of Education,
Nara 630-8528, Japan}
\author{Yasuhiro Asano}
\affiliation{Department of Applied Physics, Hokkaido University,
Sapporo 060-8628, Japan}
\author{Satoshi Tanda}
\affiliation{Department of Applied Physics, Hokkaido University,
Sapporo 060-8628, Japan}

\date{\today}

\begin{abstract}
We investigate the spacetime of a slowly rotating black hole 
in the Chern-Simons modified gravity. 
The long range feature of frame-dragging effect under the 
Chern-Simon gravity well explains the flat rotation 
curves of galaxies which is a central evidence of dark matter.
Our solution provides a different scenario of rotating space from 
G\"odel's solution.
\end{abstract}


\maketitle

{\it Introduction}.---There are three fundamental unsolved 
issues in the theory of gravity:
quantization of gravity, dark energy, and dark matter.
The string theory \cite{polch} is a promising 
candidate for a consistent quantum theory of gravity.
Many attempts in quantizing gravity, however, 
have not been successful. 
In astronomy/astrophysics, a number of observations suggest 
the existence of dark energy \cite{riess,bennett}
and dark matter \cite{bennett,dm}. 
Although many surveys of astrophysical 
objects have been conducted \cite{dm2}, 
it has not yet been revealed what is dark matter.
For instance, the flat rotation curves of galaxies \cite{rubin} 
have been considered to be a robust evidence of dark matter. 
The velocity $v$ of a star orbiting around the center of a galaxy becomes 
a constant at a certain distance $r$ far from the center. 
While the Newtonian gravity 
yields a relation $v\propto 1/\sqrt{r}$. 
At present, we usually attribute the discrepancy to dark matter. 
However it may be still possible to explain the phenomenon based on 
a theory without dark matter.
In this paper, we propose a model to solve this discrepancy in the 
framework of the Chern-Simons (CS) gravity.

The CS action 
(see Eq.~(\ref{eq:action})) is a universal entity obtained 
from an effective action in the string theory \cite{polch,seck}.
In fact, the action also appears 
in condensed matter physics such as quantum Hall effect 
\cite{jackiw,matsuyama}. 
Deser {\it et al.}~\cite{djt} originally constructed 
a theory of
CS gravity in (2+1) dimensions and 
recently Jackiw \& Pi extended it to (3+1) dimensions \cite{jp}. 
Several important consequences of the CS action have been 
discussed in theories of gravity \cite{ay,kmt,gh,seck}. 
In connection with this study, we summarize two important features of the 
CS gravity as follows.
First the Schwarzschild solution satisfies 
the field equation of CS gravity, which 
indicates that the CS theory meets a requirement of 
classical tests for general relativity \cite{will}. 
Second, generally speaking, the CS term enhances angular 
velocity or angular momentum.
In other words, the CS term rather modifies the gravitomagnetic part of the 
gravitational field than the gravitoelectric part \cite{mcl}. 
These facts imply that CS gravity may behave as if dark matter 
exists, which motivated this study.

In this paper, we study a solution for a slowly rotating black hole 
in the CS gravity 
based on a previous work \cite{kmt}. The frame-dragging effect 
turns out to be significantly large in a region far from 
the gravitational source.
As a consequence, obtained solution explains 
the flat rotation curves of galaxy.
By analyzing the frame-dragging effect on precession of a 
spinning object, 
we also discuss a way to confirm the CS gravity in observations. 
This paper indicates a route connecting the quantum theory of gravity with 
the dark matter problem.
Hereafter, we use the geometrized units with $c=G=1$.

{\it CS gravity }.---We briefly review 
a theory of the CS modified gravity \cite{jp}.
The action in this paper is given by
the Einstein-Hilbert action and the CS one,
\begin{eqnarray}
\label{eq:action}
 I & = & \frac{1}{16\pi} \int d^4 x 
  \left( - \sqrt{-g} R + \frac{1}{4} l
   \vartheta \: ^{\ast} \!R^{\sigma\ \mu\nu}_{\ \tau} 
   R^{\tau}_{\ \sigma\mu\nu} \right) , 
\end{eqnarray}
where $g$ is the determinant of the metric, 
$R \equiv g^{\alpha\beta} R^{\lambda}_{\ \alpha\lambda\beta}$ 
is the Ricci scalar, 
$R^{\tau}_{\ \sigma\alpha\beta} \equiv \partial_{\beta} 
\Gamma^{\tau}_{\sigma\alpha} - \cdots$ is the Riemann tensor
($\Gamma^{\alpha}_{\beta\gamma}$ is the Christoffel symbols), 
$l$ is a coupling constant, and
$\vartheta$ is an external quantity.
The dual Riemann tensor density is defined by 
$^{\ast} \!R^{\tau\ \mu\nu}_{\ \sigma} \equiv \frac{1}{2} 
\varepsilon^{\mu\nu\alpha\beta} R^{\tau}_{\ \sigma\alpha\beta}$,
where $\varepsilon^{\mu\nu\alpha\beta}$ is the Levi-Civita tensor
density of weight one.
 The variation of the action with respect to $g_{\mu\nu}$
gives a field equation 
\begin{equation}
\label{eq:feq}
 G^{\mu\nu} + l C^{\mu\nu} = -8\pi T^{\mu\nu} ,
\end{equation}
where $G^{\mu\nu}$ is the Einstein tensor, $T^{\mu\nu}$ is 
the energy-momentum tensor, and $C^{\mu\nu}$ is the Cotton tensor 
defined as
\begin{eqnarray}
 C^{\mu\nu} & = & - \frac{1}{2\sqrt{-g}}
 \left[ v_{\sigma} \left( \varepsilon^{\sigma\mu\alpha\beta}
   \nabla_{\alpha} R^{\nu}_{\ \beta} + 
   \varepsilon^{\sigma\nu\alpha\beta}
   \nabla_{\alpha} R^{\mu}_{\ \beta} \right) \right.
   \nonumber \\
 && \qquad \qquad + \left. \left( \nabla_{\sigma} v_{\tau} \right)
   \left( \:\! ^{\ast} \! R^{\tau\mu\sigma\nu} 
    + \:\! ^{\ast} \! R^{\tau\nu\sigma\mu} \right)\right] .
\end{eqnarray}
Here, $v_{\mu} \equiv \partial_{\mu} \vartheta$
is called embedding vector.
In this theory, the condition 
\begin{equation}
\label{eq:cndt}
 0 = \nabla_{\mu} C^{\mu\nu} 
  = \frac{1}{8\sqrt{-g}} v^{\nu} \;\! 
    ^{\ast} \!R^{\sigma\ \mu\lambda}_{\ \tau} 
    R^{\tau}_{\ \sigma\mu\lambda}
\end{equation}
should be imposed to ensure the diffeomorphism invariance. 
This condition is derived by using 
the Bianchi identity $\nabla_{\mu} G^{\mu\nu} = 0$
and the equation of motion $\nabla_{\mu} T^{\mu\nu} = 0$.
Equations (\ref{eq:feq}) and (\ref{eq:cndt}) are basic equations 
in the CS gravity.

In the previous work \cite{kmt}, we obtained a solution of Eqs.~(\ref{eq:feq}) 
and (\ref{eq:cndt}) for a slowly rotating black hole 
by using the perturbation expansion around the Schwarzschild solution.
In this paper, we particularly concentrate our attention on 
the case of a spacelike vector
$v_\mu = \partial_{\mu} \vartheta
= \partial_{\mu} z = \partial_{\mu} \left( r\cos\theta \right)
= \left( 0, \cos\theta , -r \sin\theta , 0\right)$.
The metric is provided by 
\begin{eqnarray}
\label{eq:sps}
 ds^2 & = & - \left( 1- \frac{2M}{r} \right) dt^2
  + \left( 1- \frac{2M}{r} \right)^{-1} dr^2 \nonumber \\
 && + r^2 \left( d\theta^2 + \sin^2 \theta d\phi^2 \right) 
    - 2 r^2 \omega (r) dtd\phi ,
\end{eqnarray}
where $M$ is mass of a black hole, 
the $\theta$-dependence of the $(t\phi)$-component
is restricted by Eq.~(\ref{eq:cndt}), and 
\begin{eqnarray}
\label{eq:omega}
 \omega 
 & = & \frac{C_{1}}{r^2} \left( 1- \frac{2M}{r} \right) +
     \frac{C_{2}}{r^3} \left[ r^2 -2Mr-4M^2 \right. 
     \nonumber \\
 &&  \left. 
  + 4M (r-2M) \ln \left( r-2M \right) \right] .
\end{eqnarray}
Here, $C_{1}$ and $C_{2}$ are constants characterizing 
the rotation of a black hole and are related to a small parameter 
$\epsilon (\equiv J/Mr)$, i.e., $C_{1}, C_{2} \sim O(\epsilon )$, 
where $J$ is the angular momentum.
This solution satisfies both Eqs.~(\ref{eq:feq}) 
and (\ref{eq:cndt}) up to the first order in $\epsilon$.
Within the first order of $\epsilon$, the solution does not
depend on $l$.
The differential equation for
$\omega$ can be obtained from 
the non-vanishing components of the field equation.
In the $(t\phi)$-component, the Einstein tensor gives 
the differential equation, whereas the Cotton tensor vanishes. 
In contrast in another components, 
the Einstein tensor vanishes, and 
the Cotton tensor gives the same differential equation.
Therefore, Eq.~(\ref{eq:omega}) is 
also a solution in general relativity. 
We emphasis that the Kerr solution does not satisfy Eq.~(\ref{eq:cndt}),
which is an important difference between the Einstein gravity 
and the CS gravity.

Here we briefly mention a relation between the original CS gravity
and the string theory.
In the original CS gravity, Eq.~(\ref{eq:cndt}) is imposed 
to ensure the diffeomorphism invariance.
As shown in Ref.~\cite{seck}, $\vartheta$ in Eq.~(\ref{eq:action}) 
is not an external quantity but a dynamical variable 
in the framework of string theory. 
Therefore Eq.~(\ref{eq:cndt}) is replaced by the field equation \cite{seck}:
\begin{equation}
\label{eq:feq2}
 g^{\mu\nu} \nabla_{\mu} \nabla_{\nu} \vartheta
 = - \frac{l}{64\pi \sqrt{-g}} \: ^{\ast} \!R^{\sigma\ \mu\nu}_{\ \tau} 
   R^{\tau}_{\ \sigma\mu\nu}
\end{equation}
in the string theory.
When $l$ is a non-vanishing value and $\vartheta$ is a constant, 
Eqs.~(\ref{eq:feq}) and (\ref{eq:feq2}), respectively, 
reduce to $G^{\mu\nu} = 0$ and 
$^{\ast} \!R^{\sigma\ \mu\lambda}_{\ \tau} 
R^{\tau}_{\ \sigma\mu\lambda} = 0$ in vacuum.
We emphasis that our solution in 
Eqs.~(\ref{eq:sps}) and (\ref{eq:omega}) 
also satisfies these equations.
Thus Eqs.~(\ref{eq:sps}) and (\ref{eq:omega}) describe 
a classical field which includes effects of quantum gravity.
Furthermore, we need not to align the rotational axis with
the embedding vector in this case because of $v^{\mu}=0$.
Hereafter we consider Eq.~(\ref{eq:feq2}) rather than
Eq.~(\ref{eq:cndt}) because Eq.~(\ref{eq:feq2}) makes 
the CS theory be self-consistent.

{\it Features of solution}.---We look into a solution of the spacetime
of a slowly rotating black hole.
For this purpose, we consider a scalar invariant 
$R^{\alpha\beta\gamma\delta} R_{\alpha\beta\gamma\delta}$ 
which is useful to evaluate the radial dependence of the 
frame-dragging effect. From the metric, 
the scalar invariant at large $r$ is obtained in the form
\begin{eqnarray}
\label{eq:scalar-i}
 \lefteqn{R^{\alpha\beta\gamma\delta} R_{\alpha\beta\gamma\delta}}
   \nonumber \\
 & = & \left( R^{(0) \alpha\beta\gamma\delta} +
     R^{(1) \alpha\beta\gamma\delta} \right)
     \left( R^{(0)}_{\ \ \alpha\beta\gamma\delta} + 
     R^{(1)}_{\ \ \alpha\beta\gamma\delta}\right) \nonumber \\
 & \simeq & \frac{48M^2}{r^6} 
   - \frac{4 C_{2}^2}{r^4 \sin^4 \theta} ,
\end{eqnarray} 
where ``$(0)$'' and ``$(1)$'' denote
the zeroth and the first order in $\epsilon$, respectively.
The first term stems from the Schwarzschild solution, 
and the second term corresponds to the rotation of a black hole. 
For $\theta \neq 0,\pi$, 
the scalar invariant asymptotically reduces 
to zero at large $r$. As a consequence, 
the spacetime becomes asymptotically flat at infinity.
The singularity of the rotational axis could be avoided by 
finding nonlinear or exact solutions.
We note that the frame-dragging part proportional to $r^{-4}$ 
decays more slowly than the Schwarzschild part proportional to $r^{-6}$.
On the derivation of our metric solution 
we use the absence of the Chern-Pontryagin density 
$^{\ast} \!R^{\sigma\ \mu\nu}_{\ \tau} R^{\tau}_{\ \sigma\mu\nu} $
which is a source of gravitational anomaly \cite{ga}. 
Thus the absence of gravitational anomaly gives the long range nature of the 
frame-dragging effect.
For comparison, we recall the Kerr metric whose 
$(t\phi)$-component at large $r$ is given by
$g_{t\phi}^{(\rm K)} = - 2J_{0} \sin^2 \theta / r$, where 
$J_{0}\sim O(\epsilon)$ is the angular momentum.
 For this metric, we obtain 
\begin{equation}
 R_{\rm (K)}^{\ \ \ \alpha\beta\gamma\delta} 
   R_{{\rm (K)}\alpha\beta\gamma\delta}
 \simeq  \frac{48M^2}{r^6} 
   - \frac{144J_{0}^2}{r^8} \left( 2 + \cos 2\theta \right) .
\end{equation}
Thus the Kerr metric gives rise to the rapid decay
of the frame-dragging part because the second term is proportional to $r^{-8}$.

{\it Trajectories of a test particle}.---To show more
astrophysical consequences of the solution, we investigate 
trajectories of a test particle with mass of $m$.
Since the metric does not depend on $t$ and $\phi$, 
the $t$ and $\phi$-components of 
four-momentum $p_{\mu}$ are conserved for the particle \cite{schutz}. 
Hence, it can be assumed that $p_{0} = -m E$ and $p_{\phi} = m L$, 
where $E$ and $L$ are the energy and angular momentum of the 
particle, respectively.
We focus on trajectories in the equatorial 
plane ($\theta = \pi / 2$).
 From $g_{\mu\nu} p^{\mu} p^{\nu} = -m^2$,
we obtain
\begin{equation}
\label{eq:eom1}
 \left( \frac{dr}{d\tau} \right)^2
 = E^2 - \left( 1- \frac{2M}{r} \right) 
   \left( 1 + \frac{L^2}{r^2} \right) - 2EL \omega (r) ,
\end{equation}
where $\tau$ is a proper time. Differentiating this equation 
with respect to $\tau$, we derive
\begin{equation}
\label{eq:eom2}
 \frac{d^2 r}{d\tau^2} 
 = - \frac{1}{2} \frac{d}{dr} \left[ \left( 1- \frac{2M}{r} \right) 
   \left( 1 + \frac{L^2}{r^2} \right) \right]
    - EL \omega' (r) ,
\end{equation}
with $\omega' \equiv d\omega /dr$.
These equations can be solved perturbatively because 
the last terms on the right-hand side of Eqs.~(\ref{eq:eom1}) 
and (\ref{eq:eom2}) are small compared with the other terms, 
i.e., $\omega , \omega' \sim O(\epsilon)$.

We solve Eqs~(\ref{eq:eom1}) and (\ref{eq:eom2}) up to 
the first order in $\epsilon$ for circular orbits ($r= {\rm const.}$). 
The solutions are then given by 
\begin{eqnarray}
 E & = & \frac{r-2M}{\sqrt{r(r-3M)}} \pm \sqrt{\frac{M}{r-3M}}
   \nonumber \\
 && \times
   \left[ r \omega (r) + \frac{r^2 (r-2M)}{2(r-3M)} \omega' (r)
   \right] , \\
 L & = & \pm r \sqrt{\frac{M}{r-3M}} 
  + \frac{r^4 (r-2M)}{2\sqrt{r(r-3M)^3}} \omega' (r) .
\end{eqnarray}
Using these quantities, we can calculate
$d\phi / dt = \left( dt / d\tau \right)^{-1} d\phi / d\tau$ 
for circular orbits.
The circular velocity is then obtained as
\begin{equation}
 v = r \frac{d\phi}{dt} 
  = \pm \sqrt{\frac{M}{r}} + \left[ r \omega (r) 
    + \frac{r^2}{2} \omega' (r) \right] .
\end{equation}
This is the central result of this paper.
The first term, which is a monotonically decreasing function
proportional to $r^{-1/2}$, is coming from the Schwarzschild metric.
It is very surprising that the second term becomes a constant at large
$r$, i.e.,
\begin{equation}
\label{eq:velocity}
 v \simeq \pm \sqrt{\frac{M}{r}} + \frac{C_2}{2} .
\end{equation}
Therefore, the same feature can be expected in the rotation curves 
of galaxies.
On the other hand in the case of the Kerr solution,
the frame-dragging part is negligible at large $r$ because 
$v_{\rm (K)}\simeq \pm \sqrt{M/r} - J_{0}/r^2$ is derived.
In a galaxy, there are a bulge and a disk associated with the central
black hole. For the spherically symmetric part such as a bulge, 
an outside spacetime solution has the same form as that of our solution. 
The non-spherical part such as a disk would deform 
the solution in the direction of the rotational axis.
Thus the feature of the flat rotation curve is considered to 
remain unchanged on the equatorial plane.

At first glance, the constant circular 
velocity $v$ at infinity seems to contradict 
the asymptotically flat spacetime. This, however, is 
explained as follows.
It is impossible to cover the spatial infinity by a single Minkowski spacetime.
To do so, a number of Minkowski spacetimes are necessary. 
It is possible to consider that the spatial infinity is covered 
by several finite-size areas and that each area 
is covered by a Minkowski spacetime.
In such situation, two adjacent areas are smoothly 
connected with each other by virtue of the infinitesimal curvature 
which is given by the second term of Eq.~(\ref{eq:scalar-i}).
This situation is similar to that for a vector potential field
created by a solenoid with magnetic flux $\Phi$. 
For a certain gauge, the vector potential has the form 
$A_{i} = (\Phi /2\pi ) \left( -y/( x^2 + y^2 ) , x/(x^2+y^2) , 0 \right)$, 
where $i$ represents spatial indexes.
(This is very similar to Eq.~(\ref{eq:shift}) below.)
If the exact form $A_{i}dx^{i} \equiv da$ is considered, 
any function for $a$ cannot cover the whole space.
Due to this fact, the loop integral of $A_{i}$ can give
the non-zero value $\Phi$, i.e., 
$\oint A_{i}dx^{i} = \oint da = \Phi$.

The angular momentum of the Kerr black hole does not depend on 
choices of a 2-surface within the definition of
$J=\int \sqrt{-g} \varepsilon_{\alpha\beta\theta\phi}
\nabla^{\alpha}\psi^{\beta} d\theta d\phi / 16\pi$ \cite{wald},
where $\psi^{\mu} = (0,0,0,1)$ is the Killing vector. 
When we evaluate the angular momentum
of a black hole in Eqs.~(\ref{eq:sps}) and (\ref{eq:omega})
by using the same definition,
the angular momentum depends not only on $C_{1}$
and $C_{2}$ but also on $r$.
This means that the gravitational field also has the angular 
momentum whose degree is given by $C_{1}$ and $C_{2}$.
It should also be noted that such a situation is 
a result of $\vartheta={\rm const}.$ in Eq.~(\ref{eq:feq2}).
When $\vartheta$ is changed from a constant at large $r$, 
the frame-dragging effect would also be modified. 
In particular, when $^{\ast} \!R^{\sigma\ \mu\lambda}_{\ \tau} 
R^{\tau}_{\ \sigma\mu\lambda} \neq 0$ in Eq.~(\ref{eq:feq2}), 
the solution of Eq.~(\ref{eq:feq}) may have the same feature as that of 
the Kerr solution.

{\it Precession of spinning objects}.---We discuss a way to confirm 
the CS gravity through the precession of spinning objects.
To discuss the precession of spinning objects, 
we first adopt the isotropic form of the metric 
derived from the transformation, 
$(x,y,z) = ( \tilde{r} \sin\theta \cos\phi ,
\tilde{r} \sin\theta \sin\phi , \tilde{r} \cos\theta )$, where
$r = \tilde{r} \left( 1+ M/2\tilde{r} \right)^2$.
Then we apply the weak field approximation to discuss 
more accessible situations.
Namely the post-Newtonian approximation is applied to 
the diagonal components of metric
and the leading term at large $r$ is considered in the 
off-diagonal components $g_{ti}$.
We obtain
\begin{eqnarray}
\label{eq:wf}
 ds^2 & \simeq & - \left( 1 - 2U \right) dt^2
        + \left( 1+ 2U \right) \left( dx^2 + dy^2 + dz^2 \right) 
        \nonumber \\
 && + 2 N_{i}^{(1)} (x,y,z) dx^{i} dt,
\end{eqnarray}
where $U=M/r$, and $N_{i}^{(1)}$ is given by 
\begin{equation}
\label{eq:shift}
  N_{i}^{(1)} = \left( C_{2} \frac{y\tilde{r}}{x^2+y^2} ,  
    - C_{2} \frac{x\tilde{r}}{x^2+y^2} , 0 \right) .
\end{equation}
By using the Minkowski metric $\eta_{\mu\nu}$, we define
$h_{\mu\nu} \equiv g_{\mu\nu} - \eta_{\mu\nu}$.
It is shown that 
$h_{\mu\nu}$ satisfies the Lorentz gauge condition
$\partial_{\alpha} h_{\ \mu}^{\alpha} - \partial_{\mu} 
h^{\lambda}_{\ \lambda} /2 = 0$ under Eq.~(\ref{eq:wf}).
In a local Lorentz frame momentarily comoving with 
a spinning object, the spin obeys an equation 
$dS_{(i)} / d\tau = S_{\mu} u^{\nu} \nabla_{\nu} e^{\mu}_{(i)}$,
where $e^{\mu}_{(\alpha)}$ denotes the tetrad and  
$S_{(i)} = e^{\mu}_{(i)} S_{\mu}$ is the spin vector~\cite{will}.
The four-velocity of the spinning object
$u^{\mu} = e^{\mu}_{(0)}$ is approximated as
$u^{\mu} \simeq \left( 1 , v^{k}  \right)$.
In the weak field approximation in Eq,~(\ref{eq:wf}), 
the equation becomes 
$d\bm{S} / d\tau = \bm{\Omega} \times \bm{S}$, where
$\bm{\Omega} = - (\bm{v} \times \bm{a}) /2
+ 3 (\bm{v} \times \nabla U) /2 
- \left( \nabla \times \bm{N}^{(1)} \right) /2$ 
and $\bm{a}$ denotes the acceleration vector.
This equation seems to have the same form as that 
obtained under the Kerr solution.
The last term of $\bm{\Omega}$, however, has a different form.
We obtain
\begin{equation}
\label{eq:omega_CS}
 \bm{\Omega}_{N} 
 = \left( - \frac{C_{2}}{2} \frac{xz}{(x^2 + y^2 )\tilde{r}} ,
   - \frac{C_{2}}{2} \frac{yz}{(x^2 + y^2 )\tilde{r}} , 
   \frac{C_{2}}{2 \tilde{r}} \right) 
\end{equation}
for our solution and
\begin{equation}
\label{eq:omega_Kerr}
 \bm{\Omega}^{\rm (K)}_{N} 
 \simeq \left( \frac{3J_{0}xz}{\tilde{r}^5} ,
  \frac{3J_{0}yz}{\tilde{r}^5} , 
  \frac{J_{0} \left[ 2z^2- (x^2+y^2) \right]}{\tilde{r}^5} \right) ,
\end{equation}
for the Kerr solution.
For large $r$, Eq.(\ref{eq:omega_CS}) is proportional to $r^{-1}$, 
whereas Eq.(\ref{eq:omega_Kerr}) decays as $r^{-3}$.
This qualitative difference can be measured from 
the radial dependence of the spin precession.
For such verification, observations 
should be done for spinning objects far from a black hole.
Thus $C_2$ can be estimated from observed data 
according to Eq.~(\ref{eq:omega_CS}).

{\it Discussion}.---The whole universe given 
by Eqs.~(\ref{eq:sps}) and (\ref{eq:omega}) slightly rotates.
However, it is difficult to observe the rotation at 
infinity because the angular displacement of objects
decreases proportional to $1/r$ with increasing $r$.
The behavior of our solution at infinity is in contrast
to Go\"del's solution \cite{godel} in the Einstein theory.
Equations (\ref{eq:sps}) and (\ref{eq:omega}) also mean 
that a galaxy is affected by the frame-dragging pull of another
galaxies. Thus we should observe
the correlation between the rotational velocity
in a galaxy and the peculiar velocities of other galaxies
to confirm the frame-dragging effect.
It would be possible to estimate $C_2$  
from analysis of the correlation.
By now, unfortunately, a few studies have made on this correlation. 
The frame-dragging effect might explain another evidences of dark matter, 
i.e., large velocity dispersion of galaxies in clusters \cite{dm}. 
The velocity dispersion in interacting galaxies 
may be expected to be larger than that in Newtonian picture. 
Future investigation for cosmological density 
perturbation in this direction is necessary.

We remark the difference between our theory and
the other modified gravity theories.
The modified Newtonian dynamics (MOND) \cite{mond} certainly
describes the flat rotation curves. However,
The MOND is entirely phenomenological and not supported
by a recent experiment \cite{gundl}. Similarly,
the f(R) gravity is phenomenological in practice \cite{fr}.
On the other hand, the CS gravity is directly related
to an essential part in the string theory,
and our results are basically parameter free.
Therefore, our theory should be distinguished
from the other theories.

{\it Conclusion}.---We have investigated features 
of the spacetime endowed with a slowly
rotating black hole within the framework of the Chern-Simons (CS) gravity.
We found that the CS gravity enhances 
the gravitomagnetic part of the gravitational field far from a black
hole.
As a consequence, the velocity of a test particle in a circular orbit
surprisingly becomes a constant far from a black hole 
as it found in a galaxy's rotation curve.
Thus the result explains a robust evidence of dark matter without 
introducing realistic matter. 
Our finding indicates a possibility to solve the dark matter problem 
by a new theoretical framework of the CS gravity.
To confirm the validity of this approach, we need to explain 
another evidences of dark matter such as 
large velocity dispersion of galaxies in cluster and structure
formation in the universe.
Further investigation in this direction could bridge the gap 
between purely theoretical 
quantum gravity and more realistic astrophysical phenomena.

\begin{acknowledgments}
This work was supported in part by a Grant-in-Aid
for Scientific Research from The 21st Century COE
Program ``Topological Science and Technology''.
We thank Dr. K. Hayasaki for useful discussions.
One author (K.K.) thanks Prof. T. Futamase for stimulating comments.
Analytical calculations were performed in part 
on computers at YITP at Kyoto University.
\end{acknowledgments}


\begin{thebibliography}{99}
 \bibitem{polch}
  J. Polchinski,
  {\it String theory}
  (Cambridge University Press, Cambridge, 1998).

 \bibitem{riess}
  A. G. Riess {\it et al.},
  Astron. J. {\bf 116}, 1009 (1998)

 \bibitem{bennett}
  C. L. Bennett {\it et al}.,
  Astrophys. J. Suppl. Ser. {\bf 148}, 1 (2003)

 \bibitem{dm}
  V. Trimble,
  Annu. Rev. Astron. Astrophys. {\bf 25}, 425 (1987). 

 \bibitem{dm2}
  B. R. Oppenheimer {\it et al}.,
  Science {\bf 292}, 698 (2001);
  A. J. Romanowsky {\it et al}.,
  Science {\bf 301}, 1696 (2003);
  R. Massey {\it et al}., 
  Nature (London) {\bf 445}, 286 (2007).

 \bibitem{rubin}
  V. C. Rubin,
  Science {\bf 220}, 1339 (1983).

 \bibitem{seck}
  T. L. Smith, A. L. Erickcek, R. R. Caldwell, and M. Kamionkowski,
  Phys. Rev. D {\bf 77}, 024015 (2008).

 \bibitem{jackiw}
  R. Jackiw,
  Phys. Rev. D {\bf 29}, 2375  (1984).

 \bibitem{matsuyama}
  T. Matsuyama,
  Prog. Theor. Phys. {\bf 77}, 711 (1987).

 \bibitem{djt}
  S. Deser, R. Jackiw, and S. Templeton, 
  Annals of Physics {\bf 140}, 372 (1982).

 \bibitem{jp}
  R. Jackiw and S.-Y. Pi, 
  Phys. Rev. D {\bf 68}, 104012 (2003).

 \bibitem{ay}
  S. Alexander and N. Yunes,
  Phys. Rev. D {\bf 75}, 124022 (2007); 
  Phys. Rev. Lett. {\bf 99}, 241101 (2007).

 \bibitem{kmt}
  K. Konno, T. Matsuyama, and S. Tanda,
  Phys. Rev. D {\bf 76}, 024009  (2007).

 \bibitem{gh}
  D. Guarrera and A. J. Hariton,
  Phys. Rev. D {\bf 76}, 044011  (2007).

 \bibitem{will}
  C. M. Will,
  {\it Theory and experiment in gravitational physics}
  (Cambridge University Press, Cambridge, 1993), 2nd edition.

 \bibitem{mcl}
  K. A. Moussa, G. Clement, and C. Leygnac,
  Class. Quantum Grav. {\bf 20}, L277 (2003).

 \bibitem{ga}
  T. Kimura, 
  Prog. Theor. Phys. {\bf 42}, 1191 (1969);
  L. Alvarez-Gaum\'e and E. Witten,
  Nucl. Phys. {\bf B234}, 269 (1983).

 \bibitem{schutz}
  B. F. Schutz,
  {\it A First Course in General Relativity} 
  (Cambridge University Press, Cambridge, 1985).

 \bibitem{wald}
  R. M. Wald,
  {\it General Relativity}
  (University of Chicago Press, Chicago, 1984).

 \bibitem{godel}
  K. G\"odel, 
  Rev. Mod. Phys. {\bf 21}, 447 (1949).

 \bibitem{mond}
  M. Milgrom,
  Astrophys. J. {\bf 270}, 365 (1983);
  {\bf 270}, 371 (1983); {\bf 270}, 384 (1983).

 \bibitem{gundl}
  J. H. Gundlach {\it et al}.,
  Phys. Rev. Lett. {\bf 98}, 150801 (2007).

 \bibitem{fr}
  C. Frigerio Martins and P. Salucci, 
  Mon. Not. R. Astron. Soc. {\bf 381}, 1103 (2007).

\end{thebibliography}
\end{document}